\documentclass[mathleft, final]{an}
\usepackage{graphicx,times,bm,url,color}
\graphicspath{{./fig/}{./png/}}

\newcommand{\EQ}{\begin{equation}}
\newcommand{\EN}{\end{equation}}
\newcommand{\EQA}{\begin{eqnarray}}
\newcommand{\ENA}{\end{eqnarray}}

\newcommand{\EEq}[1]{Equation~(\ref{#1})}

\newcommand{\Eq}[1]{Eq.~(\ref{#1})}

\newcommand{\Fig}[1]{Fig.~\ref{#1}}

\newcommand{\Figs}[2]{Figs.~\ref{#1} and \ref{#2}}

\newcommand{\bra}[1]{\langle #1\rangle}

{}
{}
{}
\newcommand{\meanemf}{\overline{\cal E} {}}

{}
{}
\newcommand{\meanEMF}{\overline{\mbox{\boldmath ${\cal E}$}}{}}{}
{}
{}
{}
{}
{}
{}
{}
{}
{}
{}
{}
{}
{}
{}
\newcommand{\meanUU}{\overline{\bm{U}}}

{}

\newcommand{\meanB}{\overline{B}}

\newcommand{\meanJ}{\overline{J}}

{}

{}
{}
{}

\newcommand{\eee}{\hat{\mbox{\boldmath $e$}} {}}

\newcommand{\meanBB}{{\overline{\bm{B}}}}
\newcommand{\meanJJ}{{\overline{\bm{J}}}}

\newcommand{\kk}{\bm{k}}

\newcommand{\xx}{\bm{x}}

\newcommand{\bb}{\bm{b}}

\newcommand{\oo}{\bm{\omega}}

\newcommand{\UU}{\bm{U}}

\newcommand{\uu}{\bm{u}}

\newcommand{\ff}{\mbox{\boldmath $f$} {}}

\newcommand{\nab}{{\bm{\nabla}}}

\newcommand{\ppsi}{\mbox{\boldmath $\psi$} {}}

\newcommand{\RRRR}{\mbox{\boldmath ${\sf R}$} {}}
\newcommand{\SSSS}{\mbox{\boldmath ${\sf S}$} {}}

\newcommand{\ii}{{\rm i}}

\newcommand{\const}{{\rm const}  {}}

\def\Rm{R_{\rm m}}

\def\cs{c_{\rm s}}

\def\kf{k_{\rm f}}

\def\urms{u_{\rm rms}}

\def\etat{\eta_{\rm t}}
\def\etatz{\eta_{\rm t0}}
\def\etatz{\eta_{\rm t0}}

\def\half{{\textstyle{1\over2}}}

\def\onethird{{\textstyle{1\over3}}}

\newcommand{\yapj}[3]{ #1, {ApJ,} {#2}, #3}

\newcommand{\yapjl}[3]{ #1, {ApJ,} {#2}, #3}

\newcommand{\yan}[3]{ #1, {Astron.\ Nachr.,} {#2}, #3}

\newcommand{\yana}[3]{ #1, {A\&A,} {#2}, #3}

\newcommand{\ygafd}[3]{ #1, {Geophys.\ Astrophys.\ Fluid Dyn.,} {#2}, #3}

\newcommand{\yprl}[3]{ #1, {Phys.\ Rev.\ Lett.,} {#2}, #3}

\newcommand{\ymn}[3]{ #1, {MNRAS,} {#2}, #3}
\newcommand{\ynat}[3]{ #1, {Nature,} {#2}, #3}

\newcommand{\ypre}[3]{ #1, {Phys.\ Rev.\ E,} {#2}, #3}

\newcommand{\ybook}[3]{ #1, {#2} (#3)}

\def \fh {\tilde{{\bm f}}}

\begin{document}
\title{The contribution of kinetic helicity to turbulent magnetic diffusivity}
\titlerunning{Contribution of helicity to turbulent magnetic diffusivity}
\authorrunning{A. Brandenburg, J. Schober, \& I. Rogachevskii}
\author{
A. Brandenburg\inst{1,2,3,4}\fnmsep\thanks{Corresponding author: brandenb@nordita.org}
\and
J. Schober\inst{3}
\and
I. Rogachevskii\inst{5,1,3}
}
\institute{
Laboratory for Atmospheric and Space Physics,
  University of Colorado, Boulder, CO 80303, USA
\and
JILA and Department of Astrophysical and Planetary Sciences,
  University of Colorado, Boulder, CO 80303, USA
\and
Nordita, KTH Royal Institute of Technology
  and Stockholm University,
  10691 Stockholm, Sweden
\and
Department of Astronomy, AlbaNova University Center,
  Stockholm University, SE-10691 Stockholm, Sweden
\and
Department of Mechanical Engineering,
  Ben-Gurion University of the Negev, P.O. Box 653, Beer-Sheva
  84105, Israel
}

\date{\today,~ $ $Revision: 1.32 $ $}


\abstract{%
Using numerical simulations of forced turbulence, it is shown
that for magnetic Reynolds numbers larger than unity,
i.e., beyond the regime of quasilinear theory, the turbulent magnetic
diffusivity attains an additional
negative contribution that is quadratic in the kinetic helicity.
In particular, for large magnetic Reynolds numbers,
the turbulent magnetic diffusivity without helicity
is about twice the value with helicity.
Such a contribution was not previously anticipated,
but, as we discuss, it turns out to be important when accurate
estimates of the turbulent magnetic diffusivity are needed.
}

\maketitle

\section{Introduction}

Large-scale magnetic fields in the turbulent convection zones of
stars or in supernova-driven turbulence of the interstellar medium of
galaxies evolve according to the equations of mean-field electrodynamics
and in particular the mean-field induction equation.
This equation is similar to the original induction equation for the actual
magnetic field, which includes the fluctuations around the mean magnetic field.
The presence of turbulence leads to enhanced effective magnetic diffusion,
which is often orders of magnitude larger than the microphysical value,
although this is usually not the case in numerical simulations and no
restriction concerning this ratio will be made in this paper.
If the velocity field is helical, there is, in addition to ordinary
turbulent diffusion, also the $\alpha$ effect, which can destabilize an
initially weak large-scale magnetic field and lead to its exponential
growth.

Mathematically, the evolution of the mean magnetic field $\meanBB$
is described by the equation,
\EQ
{\partial\meanBB\over\partial t}=\nab\times\left[
\alpha\meanBB-(\etat+\eta)\mu_0\meanJJ\right],
\EN
where $\meanJJ=\nab\times\meanBB/\mu_0$ is the mean current density,
$\mu_0$ is the vacuum permeability, $\eta$ is the microphysical magnetic
diffusivity, and overbars denote spatial averaging, which we will later
specify to be horizontal averaging over two spatial coordinates $x$
and $y$.
For the purpose of this discussion, and throughout this paper, we assume
the turbulence to be isotropic; otherwise, $\alpha$ and $\etat$ would
have to be replaced by tensors.

The relative importance of turbulent diffusion to microphysical diffusion
is measured by the magnetic Reynolds number,
\EQ
\Rm=\urms/\eta\kf,
\label{Rm}
\EN
where $\urms$ is the rms velocity of the turbulence and $\kf$ is the
wavenumber of the energy-carrying eddies.
The magnetic diffusivity is inversely proportional to the electric
conductivity, so in the low conductivity limit, i.e., $\Rm\ll1$,
we have (Krause \& R\"adler 1980)
\EQ
\etat=-{1\over3\eta}\left(\overline{\ppsi^2}-\overline{\phi^2}\right)
\quad \mbox{and} \quad
\alpha=-{1\over3\eta}\overline{\ppsi\cdot\uu},
\EN
where $\uu=\nab\times\ppsi+\nab\phi$ is the turbulent velocity expressed
in terms of a vector potential $\ppsi$ and a scalar potential $\phi$.
In the following, we perform averaging over two coordinate directions.

One often considers the limiting case of incompressible turbulence,
so $\phi=0$ and $\overline{\ppsi^2}=\overline{\uu^2}/\kf^2$ as well as
$\overline{\ppsi\cdot\uu}=\overline{\oo\cdot\uu}/\kf^2$, where $\kf$
is the wavenumber of the energy-carrying eddies and $\oo=\nab\times\uu$
is the vorticity.
In that case, we can write
\EQ
\etat=\onethird\tau\overline{\uu^2}
\quad \mbox{and} \quad
\alpha=-\onethird\tau\overline{\oo\cdot\uu}
\label{alpeta}
\EN
with
\EQ
\tau=(\eta\kf^2)^{-1}
\EN
being the microphysical magnetic diffusion time based on the wavenumber $\kf$.
We reiterate that this expression applies only to isotropic conditions.
Indeed, simple anisotropic flows can be constructed, where
$\overline{\oo\cdot\uu}=0$, but $\overline{\ppsi\cdot\uu}\neq0$,
and so they do yield an $\alpha$ effect (R\"adler \& Brandenburg 2003).
Furthermore, in the compressible case, there is a negative contribution
to $\etat$, so that it can even become negative, as has been demonstrated
by R\"adler et al.\ (2011).

By contrast, in the high conductivity limit, $\Rm\gg1$,
\Eq{alpeta} still applies (Krause \& R\"adler 1980), but now
\EQ
\tau\approx(\urms\kf)^{-1}\quad\mbox{($\Rm\gg1$)}
\EN
being the correlation time.
This was also confirmed numerically using the test-field method
(Sur et al.\ 2008), although our new results discussed below will
show a slight twist to the $\Rm$-dependence of their result.

\EEq{alpeta} is also motivated by dimensional arguments.
In particular, since $\alpha$ is a pseudoscalar, it is clear that
in the present case, where the only pseudoscalar in the system is
$\overline{\oo\cdot\uu}$, there can be no other contribution to $\alpha$.
This is, however, not the case for $\etat$, which is just an ordinary scalar.
Thus, in the present case, there may well be an additional contribution
proportional to $(\overline{\oo\cdot\uu})^2$, for example.
The purpose of this paper is to show that this is indeed the case.

A particularly useful diagnostics is the ratio $\etat/\alpha$,
because it is expected to be independent of $\tau$ and equal to
$\overline{\uu^2}/\overline{\oo\cdot\uu}$
in the limit of small magnetic Reynolds numbers, where \Eq{alpeta}
is obeyed exactly.
In this paper, we shall confirm that this is indeed the case when
$\Rm\ll1$, but we find a departure from this simple result as $\Rm$
is increased.
We shall use the test-field method (Schrinner et al.\ 2005, 2007), which
has been highly successful in measuring turbulent transport coefficients
in isotropic turbulence (Sur et al.\ 2008, Brandenburg et al.\ 2008b),
shear flow turbulence (Brandenburg 2005, Brandenburg et al.\ 2008a,
Gressel et al.\ 2008, Gressel 2010, Madarassy \& Brandenburg 2010),
as well as magnetically quenched turbulence (Brandenburg et al.\ 2008c,
Karak et al.\ 2014).

\section{Test field method in turbulence simulations}

As in a number of previous cases (e.g., Brandenburg 2001), we reconsider
isotropically forced turbulence either with or without helicity using an
isothermal equation of state.
Since the magnetic field is assumed to be weak, there is no backreaction
of the magnetic field on the flow.
Furthermore, instead of solving for the magnetic field, we just solve
for the fluctuations of the magnetic field that arise from a set of
given test fields.
This equation is given by
\EQA
{\partial\bb^{\rm T}\over\partial t}&=&\nab\times\left(
\uu\times\meanBB^{\rm T}+\meanUU\times\bb^{\rm T}+\uu\times\bb^{\rm T}
-\overline{\uu\times\bb^{\rm T}}\right)
\nonumber \\
&&+\eta\nabla^2\bb^{\rm T}.
\ENA
Here, $\meanUU+\uu\equiv\UU$ is the time-dependent flow, which we take
to be the solution to the momentum and continuity equations with constant
sound speed $\cs$, a random forcing function $\ff$, density $\rho$, and
the traceless rate of strain tensor ${\sf S}_{ij}=\half(U_{i,j}+U_{j,i})
-\onethird\delta_{ij}\nab\cdot\UU$ (commas denote partial differentiation),
\EQ
{\partial\UU\over\partial t}=-\UU\cdot\nab\UU-\cs^2\nab\ln\rho
+{1\over\rho}\nab\cdot(2\nu\rho\SSSS)+\ff,
\EN
\EQ
{\partial\ln\rho\over\partial t}=-\UU\cdot\nab\ln\rho-\nab\cdot\UU.
\EN
The following four test fields, $\meanBB^{\rm T}$, are used:
\EQ
\pmatrix{\cos k_1z\cr0\cr0}\!,\,
\pmatrix{\sin k_1z\cr0\cr0}\!,\,
\pmatrix{0\cr\cos k_1z\cr0}\!,\,
\pmatrix{0\cr\sin k_1z\cr0}\!.\,
\EN
For each $\meanBB^{\rm T}$, the solutions $\bb^{\rm T}$
allow us to compute the mean electromotive force,
$\meanEMF^{\rm T}=\overline{\uu\times\bb^{\rm T}}$, and relate it to
$\meanBB^{\rm T}$ and $\mu_0\meanJJ^{\rm T}\equiv\nab\times\meanBB^{\rm T}$ via
\EQ
\meanemf_i=\alpha_{ij}\meanB_j^{\rm T}-\eta_{ij}\mu_0\meanJ_j^{\rm T}.
\EN
The four independent test fields constitute eight scalar equations for
the $x$ and $y$ components of $\meanemf_i$ with $i=1$ and 2, that can
be solved for the eight unknown relevant components of $\alpha_{ij}$
and $\eta_{ij}$ with $i,j=1,2$.
The $i=3$ component does not enter, because we use averaging over $x$
and $y$, so $\meanB_3=\const=0$ owing to $\nab\cdot\meanBB=0$ and the
absence of a uniform imposed field.

For isotropically forced turbulence, we expect
$\alpha_{12}=\alpha_{21}=\eta_{12}=\eta_{21}=0$,
$\alpha_{11}=\alpha_{22}=\alpha$, and $\eta_{11}=\eta_{22}=\etat$.
This is, however, only true in a statistical sense, and since
$\alpha$ and $\etat$ are still functions of $z$ and $t$,
we must average over these two coordinates, so we compute
\EQ
\alpha=\half\bra{\alpha_{11}+\alpha_{22}}_{zt},\quad
\etat=\half\bra{\eta_{11}+\eta_{22}}_{zt},\quad
\EN
where $\bra{\cdot}_{zt}$ denotes averaging over $z$ and $t$.

We use the forcing function $\ff$ that consists of random, white-in-time,
plane waves with a certain average wavenumber $\kf$ (Brandenburg 2001),
\begin{equation}
\ff(\xx,t)={\rm Re}\{N\fh(\kk,t)\exp[\ii\kk\cdot\xx+\ii\phi]\},
\end{equation}
where $\xx$ is the position vector.
We choose $N=f_0 \sqrt{\cs^3 |\kk|}$, where $f_0$ is a nondimensional
forcing amplitude.
At each timestep, we select randomly the phase $-\pi<\phi\le\pi$ and the
wavevector $\kk$ from many possible discrete wavevectors in a certain
range around a given value of $\kf$.
The Fourier amplitudes,
\begin{equation}
\fh({\kk})=\RRRR\cdot\fh({\kk})^{\rm(nohel)}\quad\mbox{with}\;\;
{\sf R}_{ij}={\delta_{ij}-\ii\sigma\epsilon_{ijk}\hat{k}
\over\sqrt{1+\sigma^2}},\;
\end{equation}
where the parameter $\sigma$ characterizes the fractional helicity of $\ff$, and
\begin{equation}
\fh({\kk})^{\rm(nohel)}=
\left(\kk\times\eee\right)/\sqrt{\kk^2-(\kk\cdot\eee)^2}
\label{nohel_forcing}
\end{equation}
is a nonhelical forcing function.
Here, $\eee$ is an arbitrary unit vector not aligned with $\kk$,
$\hat{\kk}$ is the unit vector along $\kk$, and $|\fh|^2=1$.

We will consider both $\sigma=0$ and $\sigma=1$,
corresponding to nonhelical and maximally helical cases.
We vary $\Rm$, defined in \Eq{Rm},
by changing $\eta$ while keeping $\nu=\eta$ in all cases.
We use the {\sc Pencil Code}\footnote{\url{https://github.com/pencil-code}}
with a numerical resolution of up to $288^3$ meshpoints in the case
with $\Rm\approx120$, which is the largest value considered here.

\section{Results}

\subsection{Dependence of $\alpha$ and $\etat$ on $\Rm$}

\begin{figure}[h!]\begin{center}
\includegraphics[width=\columnwidth]{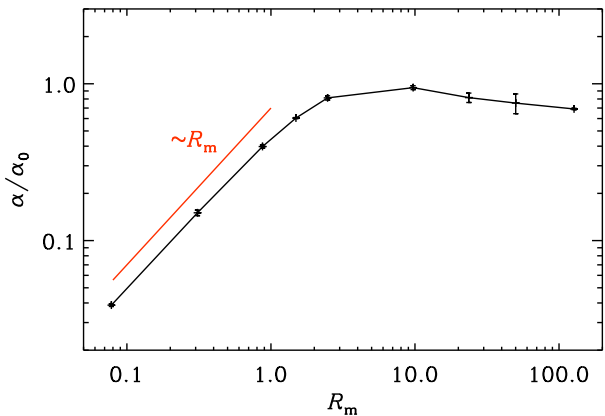}
\end{center}\caption[]{
Dependence of $\alpha$ on $\Rm$ for the models with maximum helicity.
}\label{alpha}\end{figure}

\begin{figure}[h!]\begin{center}
\includegraphics[width=\columnwidth]{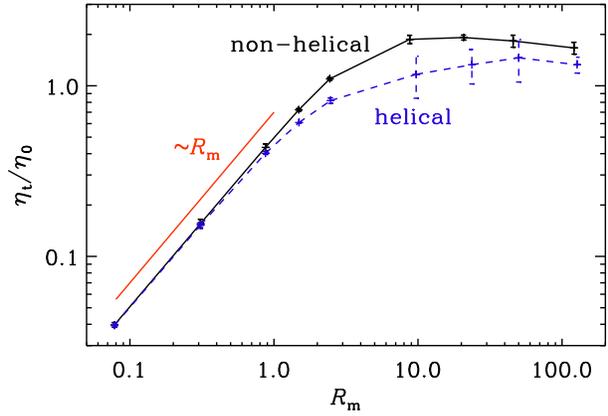}
\end{center}\caption[]{
Dependence of $\etat$ on $\Rm$ for models with maximum helicity
(dashed blue) and with zero helicity (solid black).
}\label{etat}\end{figure}

As theoretically expected (Moffatt 1978, Krause \& R\"adler 1980),
and previously demonstrated using the test-field method (Sur et al.\ 2008),
$\alpha$ and $\eta$ increase linearly with $\Rm$ for $\Rm<1$;
see \Figs{alpha}{etat} for nonhelical and helical cases.
Here, error bars have been evaluated as the maximum departure
from the averages for any one third of the full time series.
In the helical case,
both $\alpha$ and $\eta$ saturate around unity, but in the non-helical
case, $\eta$ overshoots the helical value by almost a factor of two;
see \Fig{etat}.

\subsection{Ratio of $\alpha$ to $\etat$}

\begin{figure}[h!]\begin{center}
\includegraphics[width=\columnwidth]{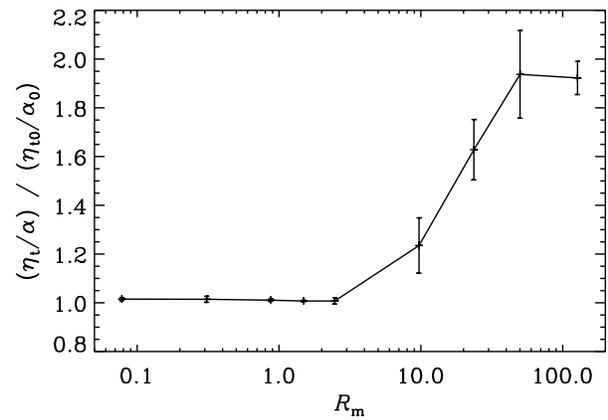}
\end{center}\caption[]{
Ratio of $\etat/\alpha$.
}\label{ratio}\end{figure}

In \Fig{ratio}, we plot the ratio $\etat/\alpha$, normalized by
$\etatz/\alpha_0$, where $\alpha_0=-\urms/3$ and $\etatz=\urms/3\kf$.
The minus sign in our expression for $\alpha_0$ takes into account
that we are forcing with positive helicity, which then leads to a
negative $\alpha$ effect (Moffatt 1978, Krause \& R\"adler 1980).
For small values of $\Rm$, this ratio is unity, but it reaches a value
of about two when $\Rm\approx50$.

\begin{figure}[h!]\begin{center}
\includegraphics[width=\columnwidth]{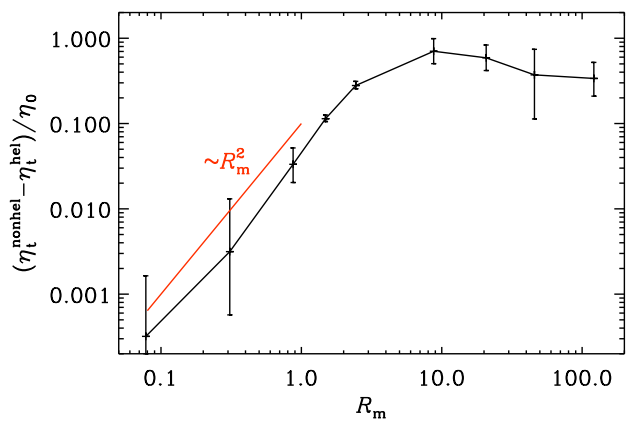}
\end{center}\caption[]{
$\Rm$ dependence of the difference between $\etat$ for models with
zero helicity an maximum helicity.
}\label{detat}\end{figure}

\subsection{Difference between nonhelical and helical cases}

It turns out that the difference between $\etat$ in the
nonhelical and helical cases increases quadratically in $\Rm$; see \Fig{detat}.
This shows first of all that the difference vanishes for small $\Rm$,
but it also suggests that there is a correction to $\etat$ due to
the presence of helicity that is not captured by the second
order correlation approximation, which is exact for $\Rm\ll1$.
It should be possible, however, to capture this effect of helicity on
$\etat$ using a higher order approximation, which has not yet been
attempted, however.

\subsection{Relation to earlier results}

A similar situation has been encountered previously in the case of the
Galloway--Proctor flow (Galloway \& Proctor 1992), where, in addition
to an $\alpha$ effect and turbulent diffusion, also a turbulent pumping
effect was found (Courvoisier et al.\ 2006).
This result was not obtained under the second order correlation
approximation (R\"adler \& Brandenburg 2009).
Using the test-field method, they showed, however, that the value of
$\gamma$, which quantifies the turbulent pumping velocity, does indeed
vanish for $\Rm\ll1$, but it was found
to increase with $\Rm$ as
$\Rm^5$; see R\"adler \& Brandenburg (2009), who interpreted this as
a higher order effect
that should be possible to capture with a six order approximation.
Our present result therefore suggests that the difference between
nonhelical and helical cases can also be described as a result of a
higher order approximation, which, in this case, would be a fourth
order approximation.

\section{Conclusions}

Our present results have demonstrated that, at least for intermediate
values of $\Rm$ in the range between 1 and 120, there is a
contribution to the usual expression for the turbulent magnetic
diffusivity $\etat=\tau\overline{\uu^2}/3$
that depends on $(\overline{\oo\cdot\uu})^2$.
This is somewhat surprising in the sense that such a result has not
previously been reported, but it is fully compatible with all known
constraints: no correction for $\Rm\ll1$ and no dependence on the sign
of $\overline{\oo\cdot\uu}$.
On the other hand, our results may still be compatible with the $\tau$
approximation in the high conductivity limit,
if the difference between
the turbulent diffusivity in the nonhelical and helical cases vanishes
for $\Rm\to\infty$.
However, our numerical results do not clearly confirm this, because our
largest value of $\Rm$ was only about 120.

There is a practically relevant application to this phenomenon, at least
in the case of forced turbulence, where its effect on the large-scale
magnetic field evolution can now be quantified to high accuracy.
A factor of nearly two in the value of $\etat$ is clearly beyond the
acceptable accuracy for this case.
This was noticed in recent studies of $\alpha$ effect and turbulent
diffusion in the presence of the chiral magnetic effect
(Schober et al.\ 2017).
Our present result therefore removes an otherwise noticeable discrepancy
relative to the theoretical predictions.
Future applications hinge obviously on the overall accuracy of analytic
approximations to particular circumstances.
In most cases, naturally driven flow turbulence will be anisotropic,
so we expect more complicated tensorial results for turbulent diffusion.

\acknowledgements
Support through the NSF Astrophysics and Astronomy Grant Program
(grant 1615100), and the Research Council of Norway (FRINATEK grant 231444),
are gratefully acknowledged.
We acknowledge the allocation of computing resources provided by the
Swedish National Allocations Committee at the Center for Parallel
Computers at the Royal Institute of Technology in Stockholm.
This work utilized the Janus supercomputer, which is supported by the
National Science Foundation (award number CNS-0821794), the University
of Colorado Boulder, the University of Colorado Denver, and the National
Center for Atmospheric Research. The Janus supercomputer is operated by
the University of Colorado Boulder.



\begin{thebibliography}{}

\bibitem{}
Brandenburg, A.\yapj{2001}{550}{824}

\bibitem{}
Brandenburg, A.\yan{2005}{326}{787}

\bibitem{}
Brandenburg, A., R\"adler, K.-H., Rheinhardt, M., \& K\"apyl\"a, P. J.\yapj{2008a}{676}{740}

\bibitem{}
Brandenburg, A., R\"adler, K.-H., \& Schrinner, M.\yana{2008b}{482}{739}

\bibitem{}
Brandenburg, A., R\"adler, K.-H., Rheinhardt, M., \& Subramanian, K.\yapjl{2008c}{687}{L49}

\bibitem{}
Courvoisier, A., Hughes, D. W., \& Tobias, S. M.\yprl{2006}{96}{034503}

\bibitem{}
Galloway, D. J., \& Proctor, M. R. E.\ynat{1992}{356}{691}

\bibitem{}
Gressel, O.\ymn{2010}{405}{41}

\bibitem{}
Gressel, O., Ziegler, U., Elstner, D., \& R\"udiger, G.\yan{2008}{329}{619}

\bibitem{}
Karak, B. B., Rheinhardt, M., Brandenburg, A., K\"apyl\"a, P. J., \& K\"apyl\"a, M. J.\yapj{2014}{795}{16}

\bibitem{}
Krause, F., \& R\"adler, K.-H.\ybook{1980}
{Mean-field Magneto\-hydro\-dy\-na\-mics and Dynamo Theory}{Oxford: Pergamon Press}

\bibitem{}
Madarassy, E. J. M., \& Brandenburg, A.\ypre{2010}{82}{016304}

\bibitem{}
Moffatt, H. K.\ybook{1978}
{Magnetic Field Generation in Electrically Conducting Fluids}
{Cambridge: Cambridge Univ.\ Press}

\bibitem{}
R\"adler, K.-H., \& Brandenburg, A.\ypre{2003}{67}{026401}

\bibitem{}
R\"adler, K.-H., \& Brandenburg, A.\ymn{2009}{393}{113}

\bibitem{}
R\"adler, K.-H., Brandenburg, A., Del Sordo, F., \& Rheinhardt, M.\ypre{2011}{84}{4}

\bibitem{}
Schober, J., Brandenburg, A., Rogachevskii, I., Boyarsky, A., Fr\"{o}hlich, J., Ruchayskiy, O., \& Kleeorin, N., 2017,
to be submitted to ApJ

\bibitem{}
Schrinner, M., R\"adler, K.-H., Schmitt, D., Rheinhardt, M., \& Christensen, U.\yan{2005}{326}{245}

\bibitem{}
Schrinner, M., R\"adler, K.-H., Schmitt, D., Rheinhardt, M., \& Christensen, U. R.\ygafd{2007}{101}{81}

\bibitem{}
Sur, S., Brandenburg, A., \& Subramanian, K.\ymn{2008}{385}{L15}

\end{thebibliography}
\end{document}